\documentclass[runningheads]{llncs}
\usepackage[T1]{fontenc}
\usepackage{graphicx}
\usepackage{url}
\usepackage{xurl}
\usepackage{booktabs}

\usepackage{color}

\usepackage{hyperref}
\hypersetup{
    colorlinks=true, 
    linkcolor=black, 
    citecolor=black, 
    filecolor=black, 
    urlcolor=black   
}

\begin{document}
\title{Across Platforms and Languages: \\Dutch Influencers and Legal Disclosures on Instagram, YouTube and TikTok}
\titlerunning{Dutch Influencers and Legal Disclosures on Instagram, YouTube and TikTok}
%
\author{Haoyang Gui\inst{1} \and
Thales Bertaglia\inst{1}\and
Catalina Goanta\inst{1} \and
Sybe de Vries\inst{1} \and 
Gerasimos Spanakis\inst{2}}
\authorrunning{Gui et al.}
%
\institute{Utrecht University, Utrecht, the Netherlands 
\email{\{h.gui,t.f.costabertaglia,e.c.goanta,s.a.devries\}@uu.nl}\\ \and
Maastricht University, Maastricht, the Netherlands
\email{jerry.spanakis@maastrichtuniversity.nl}}
\maketitle              
\begin{abstract}
Content monetization on social media fuels a growing influencer economy. Influencer marketing remains largely undisclosed or inappropriately disclosed on social media. Non-disclosure issues have become a priority for national and supranational authorities worldwide, who are starting to impose increasingly harsher sanctions on them. This paper proposes a transparent methodology for measuring whether and how influencers comply with disclosures based on legal standards. We introduce a novel distinction between disclosures that are legally sufficient (green) and legally insufficient (yellow). We apply this methodology to an original dataset reflecting the content of 150 Dutch influencers publicly registered with the Dutch Media Authority based on recently introduced registration obligations. The dataset consists of 292,315 posts and is multi-language (English and Dutch) and cross-platform (Instagram, YouTube and TikTok). We find that influencer marketing remains generally underdisclosed on social media, and that bigger influencers are not necessarily more compliant with disclosure standards. 
\end{abstract}

\section{Introduction} \label{Introduction}
Social media is undergoing fundamental changes due to the presence of users who rely on monetization, known as influencers or content creators. Influencers engage in various monetization business models, the most popular being \textit{influencer marketing} consists of brands hiring influencers to deliver advertising services in exchange for money, goods and/or services. Such ads tend to look like content rather than advertising. As a result, influencer marketing remains largely undisclosed or inappropriately disclosed on social media~\cite{ershov_effects_2020,mathur_endorsements_2018}. 

Despite an exponential interest in influencer studies across various computer science communities in the past years~\cite{ershov_effects_2020,kim_discovering_2021,zarei_characterising_2020}, the resulting body of academic work in this field has faced three main problems. \textit{First} is the problem of the evasiveness of influencer definitions and classifications.  In academic literature, influencers are defined either in terms of size~\cite{zarei_characterising_2020}, network influence~\cite{han_fitnet_2021} or based on manual curation by researchers~\cite{ershov_effects_2020}. These approaches remain unrelated to legal standards. 

\textit{Second}, not all monetized posts can be objectively identified. Thus, measuring hidden advertising generally suffers from an inherent degree of subjectivity in the perception of which content is monetized. \textit{Third}, laws worldwide establish abstract disclosure obligations but often do not include practical standards. This leads researchers to propose their own (non-legal) disclosure standards.

This study proposes a transparent methodology for measuring influencer disclosure compliance based on legal standards. We focus on the Netherlands, where both authorities and the advertising industry have been very active in setting clear disclosure standards.  We introduce a novel dataset of influencers registered with the Dutch Media Authority based on a legal registration obligation imposed in 2022 by Dutch media law~\cite{dutch_media_authority_video-uploader_2022}. We collect and analyze a multi-language and cross-platform dataset to measure and characterize the advertising disclosures by Dutch influencers. Our research makes several contributions. First, it provides a comprehensive, multi-language (English and Dutch), cross-platform (Instagram, YouTube, and TikTok) measurement of influencer marketing disclosures based on legal standards. Second, it proposes and applies an original disclosure taxonomy that distinguishes between legally sufficient (green) disclosures and legally insufficient (yellow) disclosures. Finally, it identifies a sub-dataset of affiliate marketing based on a simple and effective methodology and uses it to measure different disclosure practices across different platforms, languages, and sizes of influencers.

\section{Related Work} \label{Related Work}
Research on content monetization has primarily focused on: monetization effectiveness \cite{zak_role_2020,wibawa_role_2021,lee_why_2022}, influencer marketing strategies~\cite{alassani_product_2019}, the impact of disclosures and regulation~\cite{mathur_endorsements_2018,ershov_effects_2020,goanta_regulation_2020,james_real_2017}, and the detection of undisclosed sponsored content~\cite{zarei_characterising_2020,kim_discovering_2021,bertaglia2024instasynth}. In this context, \cite{zarei_characterising_2020} compiled a dataset of 35,000 posts and 99,000 stories from Instagram, categorizing influencers by their audience size and employing deep neural networks to distinguish between disclosed and undisclosed sponsored posts. \cite{kim_discovering_2021} compiled a large dataset of 1.6 million Instagram posts and employed network features, including brand mentions and connections between posts, to train deep learning models for detecting hidden advertisements. Additionally,~\cite{bertaglia2023closing} investigated the reliability of human annotators in detecting undisclosed ads, highlighting the implications of such inconsistencies for machine learning models. ~\cite{mathur_endorsements_2018} also applied web measurement methods and identified only 10\% AM content as disclosed out of 3,472 YouTube videos and 18,273 Pinterest pins. While these studies offer substantial insights, they exhibit a notable gap in connecting computational findings with legal standards within specific jurisdictions. 

\section{Content Monetization and Legal Disclosures} \label{Content monetization and legal disclosures}
\textbf{Influencer Marketing and Dutch law.} Based on the contractual transaction models, influencer marketing practices include \textit{Endorsements}, where money is exchanged for advertising services; \textit{Barters}, which involve goods or services being provided in return for advertising services; and \textit{Affiliate Marketing} (AM), where each sale results in a referral commission \cite{parliament_impact_2022}. In the Netherlands, media and consumer law determine applicable disclosure standards. Laws are generally vague and principle-based. However, self-regulatory organizations such as the Dutch Advertising Organization (Stichting Reclame Code) have proposed more specific rules, such as which hashtags should be clear enough for disclosure purposes. These rules are included in the Dutch Advertising Code, which theoretically must be aligned with Dutch law. In this study, we therefore focus on the more specific rules of the Dutch Advertising Organization to computationally model legally required disclosures. In parallel, the Dutch Media Authority is a state organization that has adopted specific national guidelines relating to identifying influencers. As a result, starting with 1 July 2022, Dutch influencers must register in the Video-Uploader Registry if they: (a) have more than 500k followers on Instagram, YouTube or TikTok; (b) make regular video content (at least 24 videos in the past 12 months); (c) make revenue based on the content; and (d) are registered with the Dutch Chamber of Commerce.

\textbf{A Legal Framework for Measuring Influencer Marketing.} These legal developments allow us to propose a simple and effective approach to measuring disclosures and overcome the research gaps identified above. First, the Dutch Video-Uploader Registry provides a means to identify influencers accurately based on legal criteria. This public registry, mandated by the government, includes influencers who have formalized their monetization activities through registration, offering a formal list that avoids definitional subjectivities. Second, we focus on the legal standards for disclosure as outlined in the Dutch Advertising Code. We categorize disclosures into green disclosures, which follow legal standards (e.g., specific hashtags and words in Dutch and their English translations), and yellow disclosures, which are more inconspicuous and commonly used by influencers (e.g., \#ambassador, \#partner). Lastly, we propose a method for identifying affiliate marketing (AM) as a benchmark for hidden advertising.

\section{Methodology} \label{Methodology}
\textbf{Data Collection and Cleaning.} Between August and October 2023, we collected textual data from the Dutch Video-Uploader Registry. We focus on text data as monetization disclosures remain largely communicated in writing. 209 registrations were officially made by 1 July 2023. However, this number included not only influencers but also other online media companies. We filtered out all the non-influencer accounts through annotations made by the research team, leading to a total of 150 influencers. Out of these 150 influencers, 133 are active on Instagram, 141 on YouTube, 131 on TikTok and 105 are on all three platforms. We used each of the respective platform's API (Instagram's Crowdtangle~\cite{crowdtangle_crowdtangle_2019}, YouTube Data API v3~\cite{youtube_youtube_2024} and TikTok Research API~\cite{tiktok_research_2023} to collect all the available data of the respective influencers. Due to API limitations or bugs (especially for the TikTok Research API), we could only retrieve data from 132 influencers from Instagram, 136 from YouTube and 127 from TikTok.

The collected data features a total of 300,199 posts. We used \texttt{lingua-py}~\cite{stahl_lingua-py_2024} to identify the language of each post. The resulting dataset reflects 292,315 posts recognized as either English or Dutch text. The relevant text data consists of 122,913 Instagram posts from 2011 to 2023, 128,444 YouTube video descriptions from 2007 to 2023 and 48,842 TikTok video descriptions from 2016 to 2023)

\textbf{Detecting Legal Disclosures.} We identify disclosures as follows. \textit{Green Disclosures} are legal disclosures made in compliance with the Dutch Advertising Code. The Code specifies that platform toggles must be used (e.g., the \textit{Paid partnership}) and that word disclosures must be positioned at the beginning of the text. We consider disclosure words in the first five words of each post (after tokenization and removing all punctuation) to be compliant. While all platforms in our study use the disclosure toggle, we only managed to collect disclosure toggle information from Instagram. \textit{Yellow Disclosures} are disclosures which are not legally sufficient but are still used by influencers. We identify them based on a list created using observations from the dataset and expert insights from the author team.

\textbf{Detecting Affiliate Marketing.} Based on AM textual cues, we compiled a list with the co-occurrence of these terms based on dataset observations. Our set of co-occurrence terms includes variations of these relevant words. When words co-occur in one post together, content can be categorized as AM. We checked the accuracy of this approach by manually annotating 10\% of 13,917 AM posts across the dataset, where we only found 2 false positives. 

\section{Findings}
We focus on three main research questions: First, what are the practices of Dutch influencers with respect to complying with legal standards? Second, how do Dutch micro- macro- and mega-influencers influencers disclose content on different platforms? Third, What is the engagement difference between disclosed and non-disclosed content across different platforms and influencer sizes?

\textbf{Legal Disclosure Practices.}
Overall, the amount of content voluntarily disclosed by influencers (green and yellow disclosures aggregated) shows that registered influencers only flag a marginal amount of their content as being monetized (6.53\%) and, therefore, needing disclosure. Table~\ref{tab: Distribution of disclosure and AM across the dataset} shows a general breakdown of the overall dataset and a distribution of disclosure practices and AM content across three platforms and two languages. Besides Dutch, the influencers also post content in English (43.5\%).


\begin{table*}
  \centering
  \caption{Percentage of disclosed and AM content by each platform and language}
  \resizebox{\linewidth}{!}{%
      \begin{tabular}{lcccccc}
        \toprule
         & \textbf{{Instagram English}} & \textbf{{Instagram Dutch}} &\textbf{ {Youtube English}} & \textbf{{Youtube Dutch}} & \textbf{{TikTok English}} & \textbf{{TikTok Dutch}} \\
            \midrule
            \textbf{Percentage of disclosures} & \textbf{2.834}\% & \textbf{3.708}\% & \textbf{12.851}\% & \textbf{10.146}\% & \textbf{1.120}\% & \textbf{3.013}\% \\
            Green disclosure & 1.049\% & 0.800\% & 0.011\% & 0.226\% & 0.143\% & 0.388\% \\
            \vspace{0.2 em}
            Yellow disclosure & 1.785\% & 2.908\% & 12.840\% & 9.920\% & 0.977\% & 2.624\% \\
            \textbf{Percentage of AM} & \textbf{2.841}\% & \textbf{0.460}\% & \textbf{3.182}\% & \textbf{12.777}\% & \textbf{0.278}\% & \textbf{0.141}\% \\
            Green disclosed AM & 0.076\% & 0.055\% & 0.000\% & 0.027\% & 0.004\% & 0.004\% \\
            Yellow disclosed AM & 0.164\% & 0.178\% & 1.294\% & 0.640\% & 0.046\% & 0.038\% \\
            Undisclosed AM & 2.602\% & 0.227\% & 1.888\% & 12.110\% & 0.227\% & 0.098\% \\
            \bottomrule
    \end{tabular}%
  }
  \label{tab: Distribution of disclosure and AM across the dataset}
\end{table*}

Within the disclosed content category, we note a very low usage of green disclosures in general, with YouTube English having the lowest proportion (0.011\%), where yellow disclosures are exclusively used (12.840\%). One possible explanation is that green disclosures require strict positioning, so influencers may be placing them at the end of the text. On YouTube, this is additionally problematic since text on the platform tends to be longer than Instagram or TikTok posts. Overall, this finding reveals a preference of Dutch influencers for using popular disclosure cues that do not comply with Dutch law. 

Table~\ref{tab: Distribution of disclosure and AM across the dataset} illustrates the overall amount of AM content in the dataset per platform and language (total 4.76\%), as well as how much AM is disclosed using green and yellow disclosures (total 0.49\%). While green and yellow disclosures only allow us to track disclosures that were voluntarily made by influencers, they do not reveal non-disclosed advertising. Using the AM sub-dataset as a benchmark, it is possible to identify hidden advertising as non-disclosed AM (total 4.27\%). 
Moreover, the green disclosure of AM content is meagre across all platforms and languages (even the highest is just 0.076\% on Instagram English). Except for YouTube English, most AM content from the other venues remains undisclosed (especially for YouTube Dutch, with 12.110\% of undisclosed AM content). 

Moving to disclosure positions, we calculate the position within the sentence (in \# of words) where the first disclosure word is shown. Although sentence length varies across different platforms, none of them has a median number lower than the first five words. Moreover, Instagram and YouTube have relatively different medians in English and Dutch, whereas the difference between TikTok's English and Dutch is small. 

\begin{figure}[htbp]
    \centering
    \includegraphics[width=0.4\linewidth]{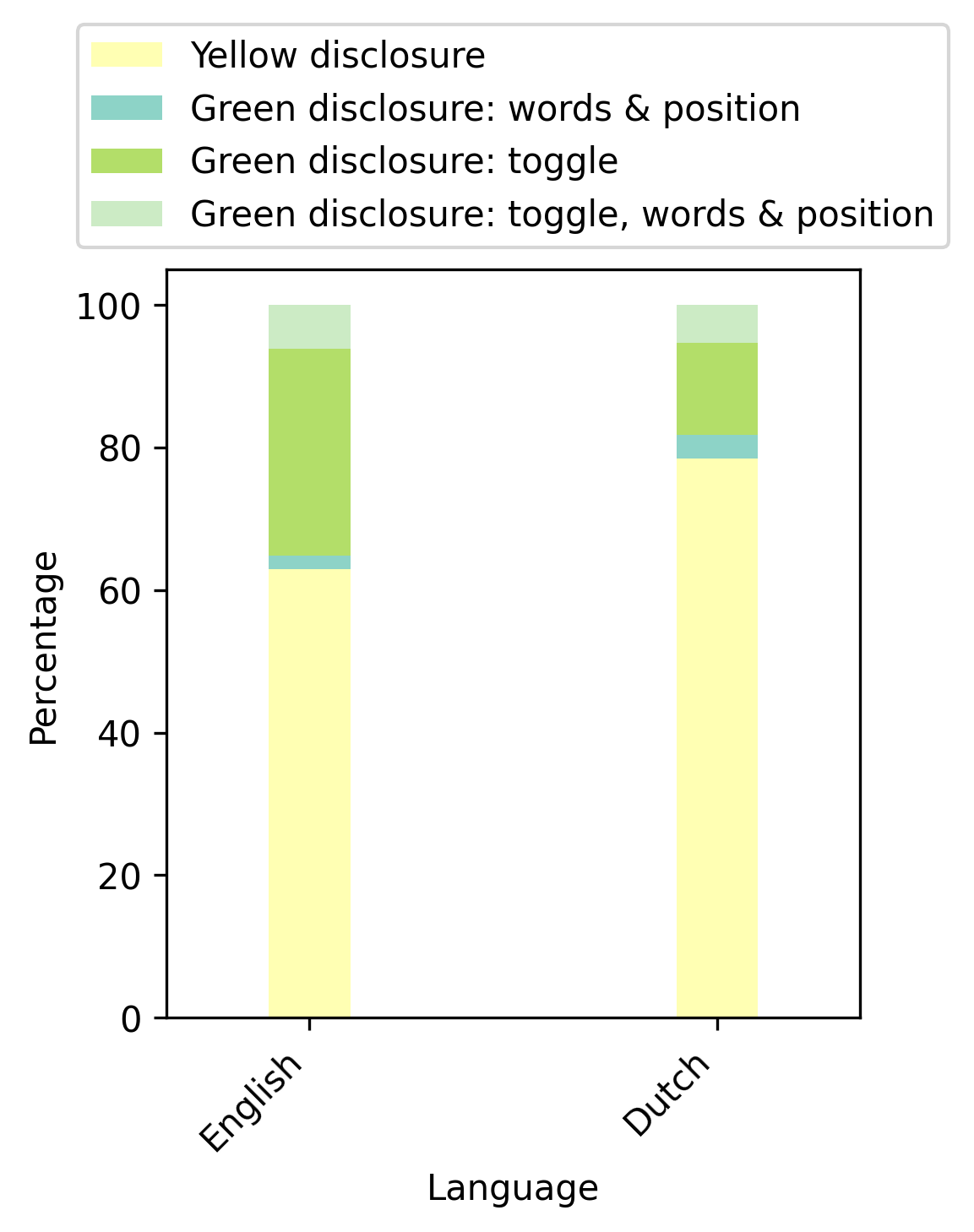}
    \caption{Instagram disclosure composition.}
    \label{fig: Instagram Disclosure Composition}
\end{figure}

While all platforms in our study use the disclosure toggle, we could only collect disclosure toggle information from Instagram. Fig.~\ref{fig: Instagram Disclosure Composition} presents the distribution of different disclosure types in Instagram data. Green disclosures are divided into three categories: \textit{Words \& position}, which refers to the right words at the beginning of the text (first five words); \textit{Toggle}, which indicates the use of the platform toggle in the platform interface; and \textit{Toggle, words, \& position}, which involves using the right words at the beginning of the text along with the platform toggle.

We find that disclosures are used insufficiently across both English (more than 60\%) and Dutch (around 80\%). Overall, there are more toggle disclosures in English than in Dutch, and for both languages, there is an insignificant amount of legal disclosures placed sufficiently early in the text. 

\textbf{Influencer Size and Disclosures.} We further investigate whether influencers with more followers disclose more monetized content and if this disclosure is legally sufficient. We determine size using the number of followers on the day when the data was collected. We divide the dataset into three audience size categories: \textit{micro-influencers} (less than 500K followers), \textit{macro-influencers} (more than 500K but less than 1M followers), and \textit{mega-influencers} (more than 1M followers). Here, we hypothesise that the bigger the following, the more professional the influencer is.
We start by looking into the general distribution of influencers by size.
Table~\ref{tab: Influencers with different sizes across the platform} shows the distribution of different sizes of influencers across platforms. 
The second and third rows in Table~\ref{tab: Influencers with different sizes across the platform} present the disclosure distribution for each platform. The results show that macro-influencers from YouTube disclose most advertising across all platforms using yellow disclosures. This corresponds with findings from Table~\ref{tab: Distribution of disclosure and AM across the dataset}, showing the higher prevalence of disclosures and AM content on YouTube compared to the other two platforms. Moreover, macro- and mega-influencers have a similar distribution of disclosure content on Instagram and TikTok, generally disclosing more than micro-influencers. 


\begin{table*}[htbp]
    \centering
    \caption{Overview of disclosure and AM by different influencer size. \# denotes the absolute number and \% the proportions of the corresponding disclosure type.}
    \label{tab: Influencers with different sizes across the platform}
    \begin{tabular}{lrrrrrrrrr}
        \toprule
        & \multicolumn{3}{c}{Instagram} & \multicolumn{3}{c}{YouTube} & \multicolumn{3}{c}{TikTok} \\
        \cmidrule(r){2-4} \cmidrule(lr){5-7} \cmidrule(l){8-10}
        & Micro & Macro & Mega & Micro & Macro & Mega & Micro & Macro & Mega \\
            \midrule
            \vspace{0.4 em}
            \# Influencers            & 73 & 35 & 24 & 65 & 43 & 28 & 58 & 35 & 32 \\
            \% Green Disclosure  & 5.51 & 7.62 & 14.90 & 0.52 & 0.71 & 0.09 & 0.72 & 5.97 & 6.17 \\
            \vspace{0.4 em}
            \% Yellow Disclosure & 18.03 & 31.05 & 22.89 & 7.01 & 73.88 & 17.78 & 10.60 & 36.11 & 40.43 \\
            \% Green AM          & 1.17 & 0.69 & 2.12 & 0.15 & 0.03 & 0.00 & 0.00 & 1.01 & 1.01 \\
            \% Yellow AM         & 2.33 & 4.24 & 4.08 & 0.75 & 4.34 & 4.35 & 0.00 & 8.08 & 12.12 \\
            \% Non-disclosed AM  & 7.68 & 5.67 & 72.02 & 2.52 & 77.87 & 9.97 & 12.12 & 41.41 & 24.24 \\
            \bottomrule
    \end{tabular}
\end{table*}

\begin{figure}
    \centering
    \includegraphics[width=0.6\linewidth]{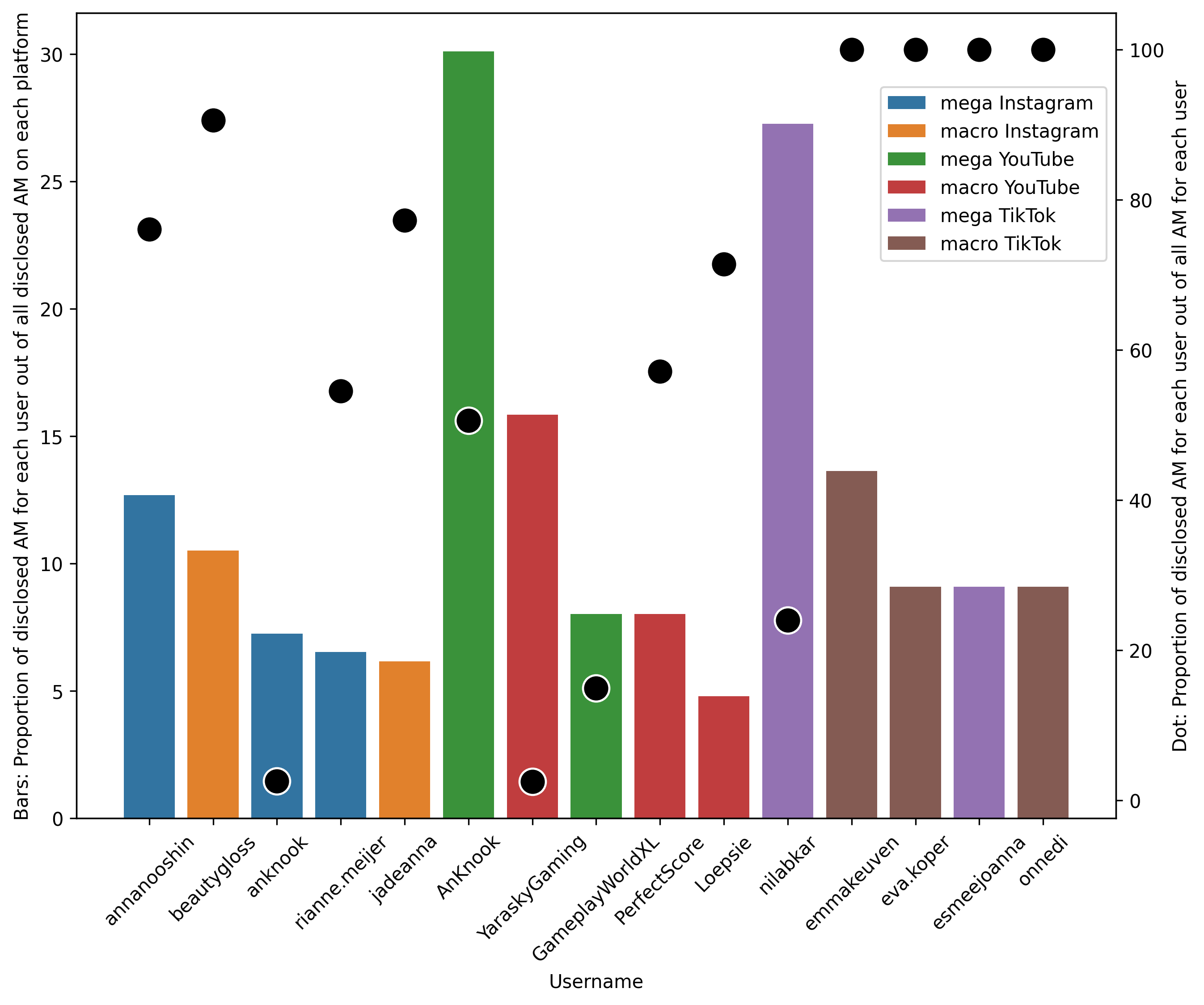}
    \caption{Top 5 accounts with the most disclosed AM on each platform and the corresponding disclosure rate for their own AM.}
    \label{fig: Top 5 accounts AM}
\end{figure}

As suggested by Table~\ref{tab: Distribution of disclosure and AM across the dataset}, almost no green disclosures are found, and the majority (around 90\%) of AM content stays undisclosed. 
These findings suggest that more disclosures originate from influencers with a large audience, whether in terms of overall disclosures or AM specifically. To further investigate this pattern, we analyze the top five influencers on each platform with the most AM disclosures. We then measure the proportion of disclosed AM among all AM created by each of the selected influencers, showing their compliance with disclosures. Fig.~\ref{fig: Top 5 accounts AM} presents the results. 

The left y-axis indicates the scale of the bars, showing the proportion of disclosed AM for each examined user out of all disclosed AM posts on each platform. As a result, the 5 influencers with the highest proportions of disclosed AM on all platforms are either macro- or mega-influencers, but none are micro-influencers. Instagram's accounts are more representative than the other two platforms, and it would not be reasonable to infer that influencers from YouTube and TikTok are more likely to disclose AM because of the skewed distribution.

Finally, the right y-axis of Fig.~\ref{fig: Top 5 accounts AM} shows the proportion of AM that is disclosed for each user (indicated by dots). Four accounts from TikTok disclose all their AM content, showing their high compliance. However, none of them have more than five AM posts in total, which makes the result not representative. In comparison, the results from Instagram and YouTube show that macro-influencers tend to disclose more AM content than mega-influencers from the same platform. These findings do not support the hypothesis that the bigger the influencers are, the more compliant they tend to be.

\begin{figure*}
    \centering
    \includegraphics[width=1\linewidth]{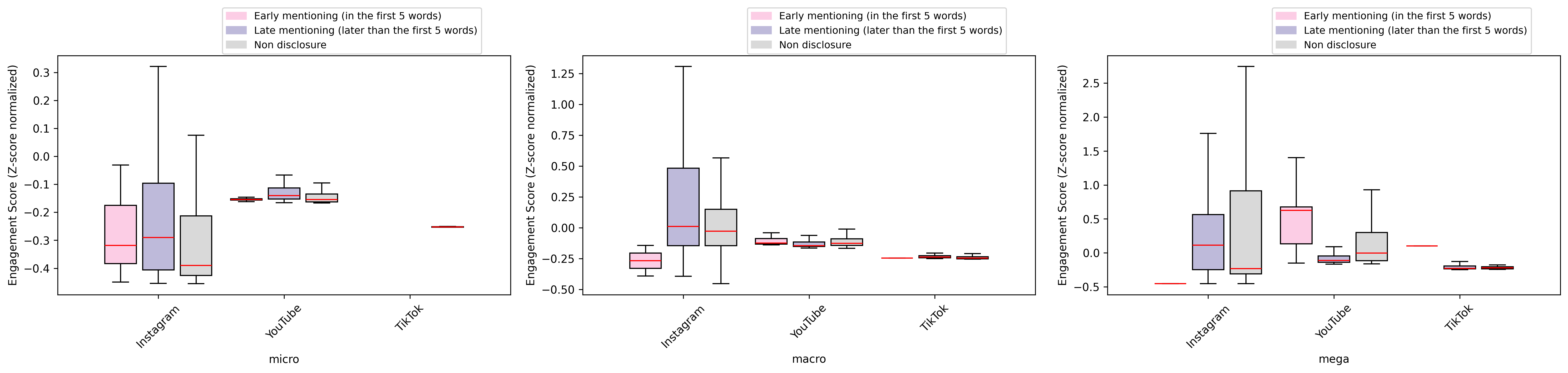}
    \caption{Box plots of engagement for AM by different disclosure word positions in each influencer category}
    \label{fig: Boxplots of engagement for AM by different disclosure word position}
\end{figure*}

\textbf{Engagement and Disclosures.} To understand how disclosures affect engagement, we conduct a series of comparative experiments on AM posts from all three platforms. For each post, we define engagement as the sum of the number of likes and comments. Fig.~\ref{fig: Boxplots of engagement for AM by different disclosure word position} shows box plots of audience engagement in AM posts for different disclosure word positions. The engagement score is normalized by the Z-score so that the results between different platforms are comparable.

This plot suggests that for TikTok few disclosures are found positioned in the first five words of the sentence for all sizes of influencers, which corresponds to the findings from Table~\ref{tab: Influencers with different sizes across the platform}. A higher median in \textit{late disclosure} can be observed on Instagram, while \textit{early disclosure} is higher than \textit{non disclosure} for mega influencers on YouTube and TikTok. Overall, these observations suggest that disclosures can benefit engagement but the results vary with the different positioning of disclosure words.

\begin{figure*}
    \centering
    \includegraphics[width=1\linewidth]{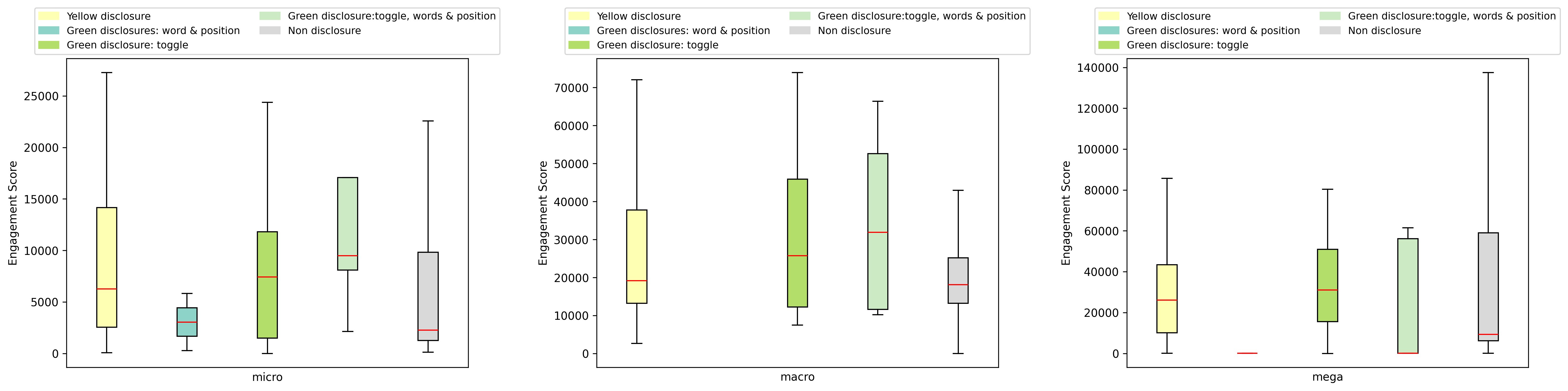}
    \caption{Engagement by different disclosure strategies of AM on Instagram}
    \label{fig: Engagement by different disclosure practices of AM on Instagram}
\end{figure*}

Lastly, we extend the experiment of the composition of different disclosures on Instagram from Fig.~\ref{fig: Instagram Disclosure Composition} and explore differences in engagement. In Fig~\ref{fig: Engagement by different disclosure practices of AM on Instagram}, ``Green disclosures: word \& position'' are only visible in micro-influencers due to its few occurrences. Except for it and ``Green disclosures: toggle, words \& position'' in mega-influencers, green disclosures tend to perform better than yellow disclosures regarding the median. The variance of green disclosures is also better than other practices in micro- and mega-influencers. Moreover, in micro- and macro-influencers, ``Green disclosure: toggle, words \& position'' also performs better regarding the median than those only using toggle. The more compliant the AM posts are, the higher engagement they tend to attract. However, findings from mega-influencers contradict this assumption, as those using both toggle and words \& position perform the worst. 

\section{Discussion and Future Research}
This paper presents granular information on how disclosures are done on social media using Dutch law as a starting point to measure legal compliance. Our analysis shows that the general volume of disclosed content is astonishingly low. The content voluntarily disclosed by influencers, whether with green or yellow disclosures, amounts to a mere 6.53\% out of the overall dataset. According to our results, in the case of affiliate marketing, only up to 10\% is disclosed, leaving 90\% of influencer marketing undisclosed. This result aligns with the low disclosure rates found in previous research on English YouTube and Pinterest affiliate marketing by~\cite{mathur_endorsements_2018}, which was, on average, around 10\%.

The growing popularity of content monetization has led to an ecosystem where influencers must be present on multiple platforms and often create content for different language audiences. It is important to understand the particularities of content creation on each of these platforms. Further research should investigate platform-specific disclosure affordances. 

\paragraph{\textbf{Limitations}}
Although we used the TikTok Research API, our data retrieval was incomplete due to API problems. We reported the issue to TikTok and used the partial data we retrieved. Data incompleteness is often seen more in earlier data points than in later ones.

\section{Acknowledgments}
This research has been supported by funding from the ERC Starting Grant HUMANads (ERC-2021-StG No 101041824).

\bibliographystyle{splncs04}
\bibliography{main}
\end{document}